\documentclass[aps,prl,reprint,groupedaddress]{revtex4-1}
\usepackage{graphicx}
\usepackage{bm}
\usepackage{float}
\usepackage{amsmath}
\usepackage{xcolor}
\usepackage{url}

\begin{document}

% \preprint{}

% Title of paper
\title{The nature of the two-peak structure in NiO valence band photoemission}


\author{Byungkyun Kang and Sangkook Choi}
\email[]{sachoi@bnl.gov}
% \homepage[]{Your web page}
% \thanks{}
% \altaffiliation{}
\affiliation{Condensed Matter Physics and Materials Science Department,
Brookhaven National Laboratory, Upton, NY 11973, USA}

% Collaboration name if desired (requires use of superscriptaddress
% option in \documentclass). \noaffiliation is required (may also be
% used with the \author command).
% \collaboration can be followed by \email, \homepage, \thanks as well.
% \collaboration{}
% \noaffiliation

\date{\today}
\begin{abstract}
  In spite of extensive studies on NiO and their accomplishments, the rich physics still raises unsolved physical problems. In particular, the nature of the two-peak structure in the valence band photoemission spectra is still controversial. By using \textit{ab initio} LQSGW+DMFT, the two-peak structure is shown to be driven by the concerted effect of antiferromagnetic ordering and intersite electron hopping. Magnetic ordering in the Ni-$e_{g}$ orbitals splits majority- and minority-spin Ni-$t_{2g}$ levels due to local Hund's coupling. Strong hybridization between O-$p$ and Ni-$e_g$, a signature of the Zhang-Rice bound state formation, boosts oxygen-mediated intersite Ni-$e_g$ orbital hopping, resulting in the enhancement of majority-spin Ni $t_{2g}$-$e_{g}$ splitting. Interestingly, these two splittings of distinct physical origins match and give rise to the observed two-peak structure in NiO. Our new understanding should be useful in designing advanced devices based on the NiO for the hole transport.

\end{abstract}
% $t_{2g}$  $e_{g}$
% \keywords{}

% \maketitle must follow title, authors, abstract, and keywords
\maketitle

% \section{Introduction}
\textit{Introduction.-} A late-transition metal monoxide NiO has been an archetypical material to explore intricate strong correlation phenomenona and propose new concepts peculiar to strongly-correlated systems. In early 1900s, NiO has been suggested, as a Mott-Hubbard insulator where its wide band gap originates from strong on-site d-d Coulomb interactions. In late 1900s,  Zaanen et al. classfied NiO as a charge-transfer insulator where its gap originated from the charge transfer between nickel and oxygen ~\cite{Zaanen_prl_1985}. In recent years, its first ionize state is regarded as a Zhang-Rice bound hole state~\cite{bala_prl_1994,kunes_prl_2007}.

Lately, the potential usage of NiO in advanced photovoltaic and spintronic devices has revived scientific and technological interests in NiO and require a deeper understanding of its electronic structure. The p-type NiO has been proposed as hole transport layers in perovskite solar cell to improve device stability against oxidation ~\cite{abzieher_adv_2019,you_nat_2016}. NiO has been used for spintronic applications to benefit from its high Neel temperature ($T_N$=525K) ~\cite{kampfrath_nat_2011,satoh_prl_2010,lin_prl_2016,machado_prl_2017}.

\begin{figure}[h]
  \centering
  \includegraphics[width=1.0\columnwidth]{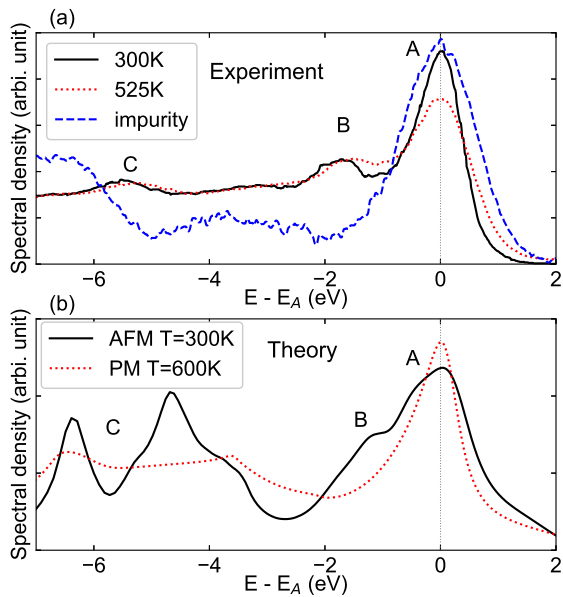}

  \caption{(a) Valence band photoemission experiments of NiO taken at 300 K (black solid line), NiO taken at Neel temperature (525 K, red dotted line), and NiO impurity in MgO (blue dashed line). The experimental spectra have been obtained from Fig. 5 and 6 in Ref.~\cite{kuo_epj_2017}. The zero of energy are set to peak A energy. (b) Total density of states of antiferromagnetic (AFM) phase (black solid line) as well as paramagnetic (PM) phase (red dotted line) of NiO within \textit{ab initio} LQSGW+DMFT. The simulation temperatures are 300 and 600K for AFM and PM phases, respectively. The zero of energy is set at the peak A energy. The first (main), second, and third peaks from the Fermi level are denoted by A, B, and C respectively on both of experimental and calculated spectra.\label{fig_1}}
\end{figure}

In spite of the extensive studies and accomplishments, the rich physics still raise unclear physical problems in NiO. In particular, the nature of the two-peak structure in angle-integrated valence-band photoemission spectra are still unclear. Fig. \ref{fig_1} (a) shows valence band photoemission spectra below the Neel temperature \cite{kuo_epj_2017}. Two peaks (A and B) are designated as main peaks in % both of X-ray core-level and
valence band photoemission spectra % of numerous experiments
and also observed in X-ray core-level photoemission~\cite{hufner_prb_1973,elp_prb_1992,alders_prb_1996,woicik_prb_2001,oh_prb_1982,taguchi_prl_2008,kuo_epj_2017,woick_prb_2018}. It has been known experimentally that non-locality and antiferromagnetic ordering are essential to give rise to the two-peak structure. In the valence band photoemission on NiO impurity embedded in MgO, B peak is strongly suppressed, implying the non-local nature of the peak B \cite{haupricht_LocalCorrelationsNonlocal_2012}. In addition, peak A dominates the line shape and peak B is suppressed at the Neel temperature, inferring its magnetic origin \cite{kuo_epj_2017}.

 % and it is enhanced by the antiferromagnetic ordering.

To understand the nature of the two-peak structure theoretically, various approaches are being pursued. One of the approaches is configuration interaction (CI) calculation of cluster models. From the single NiO$_{6}$ cluster calculation~\cite{elp_prb_1992,fujimori_prb_1984}, important low-lying many-body states have been identified including $^{4}T_{1}$ high-spin state (a hole in the minority-spin Ni-$t_{2g}$ orbitals and surrounding O-$p$ orbital), $^{2}E$ low-spin state (a hole in the majority-spin Ni-$e_g$ orbitals and surrounding O-$p$ orbital), and $^{2}T_{1}$ low-spin state (a hole in the majority-spin Ni-$t_{2g}$ orbitals and surrounding O-$p$ orbital). % By adding additional orbitals to the single NiO$_{6}$ cluster model, non-local nature of the A and B peak have been studied more seriously and
Motivated by the experimental implication, several ideas to take into account non-local and magnetic nature have been suggested and tested by expanding the size of the cluster. They succeed to reproduce the observed two-peak structure but reached to different interpretation on the nature of the two-peak structure. Taguchi et al.~\cite{taguchi_prl_2008} added additional effective orbitals, playing the role of playing the role of Zhang-Rice doublet bound state, to the single NiO$_{6}$ cluster calculation. They concluded that the two-peak structure is attributed to the effective screening orbitals playing the role of Zhang-Rice bound states with $e_g$ symmetry. Kuo et al. expanded the size of the cluster to Ni$_3$O$_{16}$ chain made of three corner-shared NiO$_6$ octahedra. They concluded that the two-peak structure is due to $^{4}T_{1}$ high-spin state and $^{2}E$ low-spin state \cite{kuo_epj_2017}.

Strong cluster model dependence on the nature of the two-peak structure has motivated the calculation of the line-shape of periodic system from \textit{first principles}. The advantage of this approach is that we can avoid issues with possible boundary effects as wells as the choice of model and its parameters. It provides density of state, which can be interpreted as a photoemission line-shape without matrix element. Various \textit{ab initio} studies have been already conducted to investigate electronic structure of NiO ~\cite{koller_prb_2011_lda,gillen_jpcm_2013_lda,trimarchi_prb_2018_lda,luen_prb_2005_gw,jiang_prb_2010_gw,das_prb_2015_gw,kunevs_prb_2007,kunes_prl_2007,miura_prb_2008,nekrasov_jetp_2012,ren_prb_2006,thunstrom_prl_2012,yin_prl_2008,lechermann_2019} but it turns out that understanding the nature of the two-peak structure from \textit{first principles} is a challenging task.
% and provide density of state, which can be interpreted as a photoemission line-shape without matrix element, ranging from density function theory ~\cite{koller_prb_2011_lda,gillen_jpcm_2013_lda}, \textit{ab initio} quasi-particle GW ~\cite{luen_prb_2005_gw,jiang_prb_2010_gw,das_prb_2015_gw} as well as LDA+DMFT ~\cite{kunevs_prb_2007,kunes_prl_2007,miura_prb_2008,nekrasov_jetp_2012,ren_prb_2006,thunstrom_prl_2012,yin_prl_2008}.
To illustrate, density functional theory doesn't display peak A and B splitting below Neel temperature. Although \textit{ab initio} quasiparticle GW provides two-peak structure below Neel temperature, but it is hard to explain the observed temperature dependence of the two-peak structure since NiO is a metal above Neel temperature in the framework. % LDA+DMFT anticipates that NiO is an insulator above Neel temperature, but it doesn't show two peak structure below Neel temperature.
Recently, self-interaction corrected LDA+DMFT reproduced two-peak structure in the paramagnetic phase and peak B is attributed to O-$p$ orbitals ~\cite{lechermann_2019}, although both peaks (A and B) are known to originate from Ni-$d$ orbitals experimentally~\cite{oh_prb_1982, haupricht_LocalCorrelationsNonlocal_2012}.

To understand the nature of the two-peak structure, we need a parameter-free framework suited for both paramagnetic (PM) and antiferromagnetically (AFM) ordered phases. One of the promising frameworks is \textit{ab initio} LQSGW+DMFT. In this parameter-free approach, non-local electron correlation is treated within \textit{ab initio} linearized quasiparticle self-consistent GW and strong local correlation is within dynamical mean-field theory. In this study, by using \textit{ab initio} LQSGW+DMFT framework, we calculated the electronic structure of both paramagnetic and antiferromagnetically ordered phases of NiO and investigated the nature of the two-peak structure.

% To get rid of finite size effect which is ineluctable in CI cluster calculations and to take into account  temperature effect on 3d valence photoemission spectra of NiO, we used parameter free LQSGW+DMFT and charge self-consistent LDA+DMFT. We shall discuss origin of peak B considering non-local correlation and temperature effect explicitly.

% The method we used in this work is based on GW+EDMFT which is a diagrammatically controlled ab-initio theory~\cite{sun_prb_2002,biermann_prl_2003,nilsson_prm_2017}, where GW method is formulated by full green function G and screened Coulomb interaction W in the full Hilbert space and local higher order diagrams are formulated within DMFT in the correlated subspace. The first order non-trivial correction is obtained within local GW approximation. We use COMSUITE package~\cite{choi_cpc_2018}, where the LQSGW+DMFT and charge self-consistent LDA+DMFT are implemented. COMSUITE mainly consists of FlawpMBPT for the GW/LDA part~\cite{kutepov_cpc_2017,kutepov_prb_2009,kutepov_prb_2012} and CTQMC for impurity solver [ ], where Coulomb interaction tensor is obtained within constrained random phase approximation (cRPA)~\cite{aryasetiawan_prb_2004}. The analytical continuation is conducted by using maximum entropy method~\cite{jarrell_pr_1996}.

% \subsection{}
% \subsubsection{}
\textit{Methods.-} All calculations are conducted using COMSUITE package~\cite{choi_cpc_2018}, where both \textit{ab initio} LQSGW+DMFT and charge self-consistent LDA+DMFT are implemented. COMSUITE is built on FlapwMBPT for the GW/LDA calculations~\cite{kutepov_cpc_2017,kutepov_prb_2009,kutepov_prb_2012}. Experimental lattice constant 4.17 \AA ~\cite{bartel_prb_1971} is adopted for the construction of two-site NiO unitcell in Rhombohedral R$\bar{3}$m space-group for both of AFM and PM simulations. % to take account into non-local correlation effect explicitly
For the AFM simulations, we constructed antiferromagnetic ordering of AFM II which is proposed as the ground state ordering for NiO~\cite{terakura_prb_1984}. The radii ($R_{MT}$) of basis of muffin-tin (MT) used in FlapwMBPT are set 2.12 bohr radius for nickel (Ni) and 1.77 bohr radius for oxygen (O). Wave functions are expanded in the MT spheres by spherical harmonics with l up to 6 for Ni and 6 for O and in the interstitial (IS) region by plane waves with the energy cutoff determined by $R_{MT,N_{i}}$ $\times$ $K_{max}$ = 6.7. Sampling the Brillouin zone is conducted using 5 $\times$ 5 $\times$ 5 k-point grid. For the GW calculation, we adopted product basis which are expanded in the MT spheres by spherical harmonics with l up to lmax=6 and in IS region $R_{MT,N_{i}}$ $\times$ $K_{max}$ = 10.0. For polarizability and self-energy calculations, all unoccupied states are taken into account. Spin-orbit coupling was not included.

For \textit{ab initio} LQSGW+DMFT calculation, five Ni-$d$ orbitals are chosen as correlated orbitals and Wannier functions for Ni-$s$, Ni-$p$ Ni-$d$, and O-$p$ orbitals are constructed in a frozen energy window of $\text{-}10 eV <E\text{-}E_f < 10eV$. For the charge self-consistent LDA+DMFT, five Ni-$d$ orbitals are considered as correlated orbitals. To define five Ni-$d$ orbitals, Wannier functions for Ni-$s$, Ni-$p$ Ni-$d$, and O-$p$ orbitals are constructed in a frozen energy window of $\text{-}10 eV <E\text{-}E_f< 10eV$ at every charge self-consistency loop. $F^0=10.0eV$, $F^2=7.8eV$ and $F^4=4.8eV$ are chosen from EDMFTF database \cite{_Kristjandatabase_} to construct Coulomb interaction tensor associated Ni-$d$ orbitals, corresponding to $U=10.0eV$ and $J=0.9eV$. For the double-counting energy, nominal double-counting scheme \cite{haule_DynamicalMeanfieldTheory_2010,haule_CovalencyTransitionmetalOxides_2014} is used with $d$ orbital occupancy of 8.0.
We tested main calculation with larger lattice constant 4.19 $\AA$ and different basis set. We did not observe appreciable changes in our main result of origin of two-peak structure.

\textit{Results.-} In Fig. \ref{fig_1} (b), the total density of states of NiO in both AFM and PM phases are compared to experimental valence band photoemission spectra. The simulation temperatures are 300K for AFM and 600K for PM phases, respectively. % Photoemission spectra are measured at 300 K for AFM phase and Neel temperature (525K) for PM phase ~\cite{kuo_epj_2017}.
Above Neel temperature, we couldn't observe the two-peak structure. Below Neel temperature, \textit{ab initio} LQSGW+DMFT predicts clear two-peak structure. This behavior is consistent with experimental observation of the enhancement of the peak B upon cooling, which is suppressed at the Neel temperature. % Here we note that the intensity of peak C is much higher in calculated spectra. The supression of the C peak in experiments might be caused by extrinsic effects as well as small photoionization cross-section of O-$p$ ~\cite{yeh_1985}, which dominates peak C. The splitting of peak C in the calculated spectra is consistent with experimental report where the splitting is attributed to multiple charge transfers~\cite{woick_prb_2018}.

\begin{figure}[h]
  \centering
  \includegraphics[width=1.0\columnwidth]{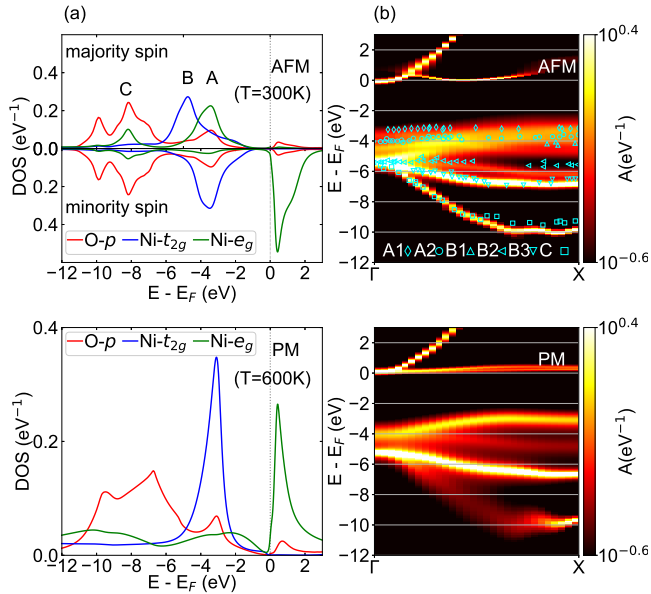}
  \caption{(a) Projected density of state of NiO in (upper panel) AFM (T=300K) and (lower panel) PM (T=600K) phases within \textit{ab intio} LQSGW+DMFT. The zero of energy is set at the chemical potential (E$_F$). The majority-(minority-)spin indicates Ni up-(down-) spin at up-(up-)spin Ni site and down-(up-)spin at down-(down-)spin Ni site. The O-$p$ is presented on both of majority- and minority-spin. (b) Momentum-resolved total spectral functions along the $\Gamma$-$X$ line in the First Brillouin zone of primitive cell of NiO in (upper panel) AFM (T=300K) and (lower panel) PM (T=600K) phases within \textit{ab intio} LQSGW+DMFT. The experimental data with six symbols has been obtained from Fig. 6 in Ref.~\cite{shen1991electronic}. % The triangle up symbol (B1) is re-classified separating from circle one.
    The experiment data are aligned with calculated valence band edge.  % Three characterized bands are denoted by A, B and C which are comprised of majority-spin $e_{g}$ + O $p$ and minority-spin $t_{2g}$, majority-spin $t_{2g}$ and mainly O $p$, respectively.
    \label{fig_2}}
\end{figure}

To identify orbital characters of these peaks, we calculated the orbital-resolved density of states for both PM and AFM phases within \textit{ab initio} LQSGW+DMFT, as shown in Fig. \ref{fig_2} (a). In the PM phase, the single peak is dominated by Ni-$t_{2g}$ orbitals and Ni-$e_{g}$ as well as O-$p$ also contribute to the single peak. % The hybridization of Ni-$e_g$ and O-$p$ orbitals at valence band edge is attributed to the formation of Zhang-Rice bound hole \cite{bala_prl_1994,kunes_prl_2007}.
In AFM phase with two-peak structure, the peak A is attributed to majority-spin Ni-$e_g$ orbitals, O-$p$ orbitals and minority-spin Ni-$t_{2g}$ orbitals. The peak B is dominated by majority-spin Ni-$t_{2g}$. Here, the majority-(minority-) spin indicates Ni up-(down-) spin at up-(up-)spin Ni site and down-(up-)spin at down-(down-)spin Ni site. Interestingly, the Ni-$t_{2g}$ magnetic splitting matches majority-spin Ni $t_{2g}$-$e_{g}$ splitting, making A and B peaks distinct. Peak C at lower energy side is mainly comprised of O-$p$.

In order to validate its momentum vector dependence, momentum-resolved spectral functions ($A(\mathbf{k},\omega)$) is calculated and compared with angle-resolved photoemission data~\cite{shen1991electronic}. For the comparison with the experiments, spectral function is unfolded into the first Brillouin zone of the primitive cell of the paramagnetic phase in the following way.
\begin{equation}
  \begin{split}
    A(\mathbf{k},\omega)=-\frac{1}{\pi}\sum_{m,M,M',\mathbf{K}}Im\langle W_{m\mathbf{k}}|W_{M\mathbf{K}}\rangle G_{MM'}(\mathbf{K},\omega)\langle W_{M'\mathbf{K}}|W_{m\mathbf{k}}\rangle ,
  \end{split}
\end{equation}
where $|W_{m\mathbf{k}}\rangle$ ($|W_{M\mathbf{K}}\rangle$) is the bloch sum of the m$_{th}$(M$_{th}$) Wannier function with the crystal momentum vector $\mathbf{k}$ ($\mathbf{K}$) in the primitive (super) cell. $G_{MM'}(\mathbf{K},\omega)$ is the Green's function in the AFM supercell. % This value is proportional to the photoelectron DC current without electron-light matrix element effect ~\cite{fujikawa2002many}.

As shown in Fig. \ref{fig_2} (b), the calculated spectral functions in the AFM phase is in an excellent agreement with experimental data measured at room temperature along the $\Gamma$-$X$ line (parallel to $\hat{x}$ direction) of the paramagnetic unitcell ~\cite{shen1991electronic}. The six bands below Fermi level in the AFM phase are identified by comparing the experimental data with orbital-decomposed spectral functions. The flat A1 and A2 bands contribute to the peak A and originate from majority-spin Ni-$e_g$, minority-spin Ni-$t_{2g}$ and O-$p$ orbitals. The band B1, B2, and B3 contribute to the peak B. B1 is from majority-spin Ni-$d_{xy}$ and $d_{xz}$ orbitals. B2 is a $d_{yz}$ dispersion-less band which has no net hopping to O-$p$ orbitals. Dispersive B3 band is attributed to the $\pi$ bonding between O-$p$ orbitals and Ni-$d_{xy}$ as well as Ni-$d_{xz}$. Dispersive C band for peak C is characterized by O-$p$ orbitals which has sizeable $\sigma$ bonding with Ni-$e_{g}$ orbitals. % For the comparison with experiments along the the $\Gamma$-$K$ line, please see the supplemental materials[ref].

\begin{figure}[h]
  \centering
  \includegraphics[width=1.0\columnwidth]{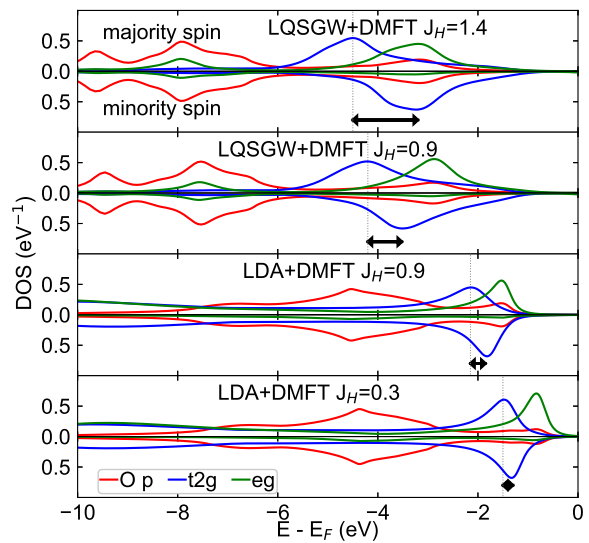}
  \caption{Calculated orbital resolved spectra intensity with various $J_H$. The energy difference between majority-spin Ni-$t_{2g}$ and minority-spin Ni-$t_{2g}$ denoted by horizontal arrows are 1.25, 0.72, 0.31 and 0.15 eV within \textit{ab initio} LQSGW+DMFT ($J_H$=1.4 eV, the value from cRPA), LQSGW+DMFT with $J_H$=0.9 eV, LDA+DMFT with $J_H$=0.9 eV and LDA+DMFT with $J_H$=0.3eV, respectively.  All spectra are calculated at 300K. The zero of energy is set at the chemical potential E$_F$.\label{fig_3}  }
\end{figure}

% \textbf{Kuo et al. suggested that weak peak B in the PM state is caused by reducing number of AFM pairing from six to three in AFM state ~\cite{kuo_epj_2017}. However our simulations for PM phase, where the impurity electron interact dynamic mean-field bath, does not include spontaneous magnetic interaction resulting in single peak in PM phase.}
\begin{figure}[h]
  \centering
  \includegraphics[width=1.0\columnwidth]{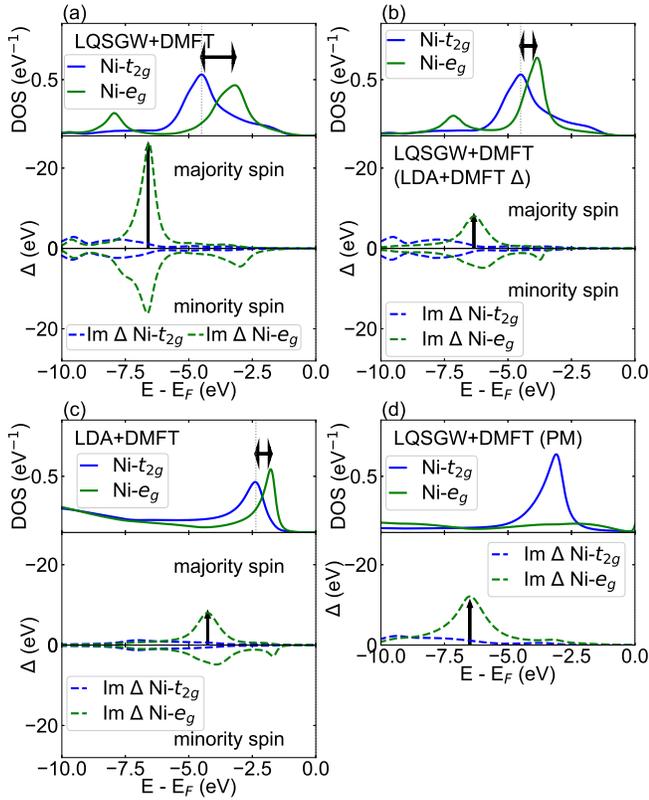}
  \caption{Calculated projected density of states (PDOS) and the imaginary part of the hybridization function of NiO in, (a) AFM phase within \textit{ab initio} LQSGW+DMFT, (b) \textit{ab initio} LQSGW+DMFT with the hybridization of Ni-$e_{g}$ from LDA+DMFT and (c) AFM phase within LDA+DMFT and (d) PM phase within \textit{ab initio} LQSGW+DMFT. For the PDOS, only majority-spin spectra of Ni-$e_{g}$ and Ni-$t_{2g}$ are presented in AFM simulations. The maximum imaginary part of hybridization of Ni-$e_{g}$ is denoted by vertical arrow. The energy difference between majority-spin Ni-$e_{g}$ and Ni-$t_{2g}$ is denoted by horizontal arrow. The simulation temperatures are 300 and 600K for AFM and PM phases, respectively. The zero of energy is set at the chemical potential E$_F$.\label{fig_4}}
\end{figure}

The concerted effect of the antiferromagnetic ordering and intersite electron hopping realizes the two-peak structure below Neel temperature. Magnetic ordering in the Ni-$e_{g}$ orbitals splits majority- and minority-spin Ni-$t_{2g}$ levels due to local Hund's coupling. The most dominant orbital characters of A and B peaks are minority-spin Ni-$t_{2g}$ orbitals and majority Ni-$t_{2g}$ orbitals, respectively. The A and B peak splitting is in the order of $J_H$. Fig. \ref{fig_3} shows the projected density of states and its dependence on J$_H$. One can see that the splitting between majority- and minority-spin Ni-$t_{2g}$ states decreases as J$_H$ decreases within both charge self-consistent LDA+DMFT as well as \textit{ab initio} LQSGW+DMFT. This magnetic splitting is close to the magnetic exchange-splitting associated with the Ni-$t_{2g}$ orbitals, which is 1.5, 1.06, 0.89, and 0.35 eV within LQSGW+DMFT ($J_H$=1.4, which is from cRPA), LQSGW+DMFT with $J_H=0.9 eV$, LDA+DMFT with $J_H=0.9 eV$ and LDA+DMFT with $J_H=0.3 eV$, respectively.

O-mediated electron hopping between Ni-$e_g$ orbitals sitting at two opposite-spin Ni atoms governs the majority-spin Ni $t_{2g}$-$e_{g}$ splitting. Fig. \ref{fig_4} (a) illustrates hybridization functions of Ni-$e_g$ orbitals within \textit{ab initio} LQSGW+DMFT. Hybridization functions of Ni-$e_g$ orbitals for both spin channels have a strong divergence at an energy where O-$p$ density of state dominates, showing strong electron hopping between Ni-$e_g$ orbitals and adjacent O-p orbitals regardless of its spin, which is a signature of Zhang-Rice bound state. Within DMFT framework, divergent hybridization pushes electron density away from the energy where hybridization diverges. In AFM NiO, this divergence pushes majority spin Ni-$e_g$ electron density toward lower-excitation energy. % Interstingly, the majority-spin Ni $t_{2g}$-$e_{g}$ splitting matches the splitting between majorigy-spin Ni-$t_{2g}$ orbitals and minority-spin Ni-$e_g$ orbitals and makes A and B peaks more distinct.
This can be demonstrated more evidently if we make a comparison with charge self-consistent LDA+DMFT results. As shown in Fig. \ref{fig_4} (c), within LDA+DMFT, Ni-$e_g$ hybridization is divergent but its divergence is relatively weak compared to \textit{ab initio} LQSGW+DMFT, resulting in smaller majority-spin Ni-$t_{2g}$-$e_{g}$ splitting. To confirm this idea, we constructed an auxiliary local Green's function ($G_{aux}(\omega)$): $G_{aux}(\omega)=(G_{LQSGW+DMFT}^{-1}(\omega)+\Delta_{Ni-{e_g}, LQSGW+DMFT}(\omega)-\Delta_{Ni-{e_g}, LDA+DMFT}(\omega))^{-1}$. Here $G_{LQSGW+DMFT}$ is \textit{ab initio} LQSGW+DMFT local Greens' function, $\Delta_{Ni_{e_g}, LQSGW+DMFT}$ is \textit{ab initio} LQSGW+DMFT hybridization function for Ni-$e_g$ orbitals, and $\Delta_{LDA+DMFT}$ is charge self-consistent LDA+DMFT hybridization function for Ni-$e_g$ orbitals. As shown in Fig. \ref{fig_4} (b), the smaller Ni-$e_g$ hybridization makes the splitting between majority-spin Ni-$t_{2g}$ and majority-spin Ni-$e_g$ smaller.

% This strong hopping between Ni-$e_g$ and O-$p$ orbitals can be attributed to the formation of the Zhang-Rice bound states.

% The second is the enhanced intersite electron hopping between Ni-$e_g$ orbitals sitting at two opposite-spin Ni atoms via O-$p$ orbitals. This results in the energy splitting between majorigy-spin Ni-$t_{2g}$ orbitals and majorigy-spin Ni-$e_g$ orbitals. Interestingly this splitting matches with the exchange splitting of Ni-$t_2g$ orbitals and makes A and B peaks more distinct. (\textbf{pdos oxygen x, y, z}).

Both majority- and minority-spin Ni-$e_g$ orbitals has strong divergence at the energy where O-$p$ density of state dominates. This means electron hops between Ni-$e_g$ orbitals sitting on two opposite-spin Ni atoms via O-$p$ orbitals. In the AFM phase, up-spin electrons in the occupied $e_g$ orbitals in up-spin Ni atom can hope to unoccupied up-spin $e_g$ orbital in down-spin Ni atom via O-$p$ orbital. At the same time, down-spin electrons in the occupied $e_g$ orbitals in down-spin Ni atom can hope to unoccupied down-spin $e_g$ orbital in up-spin Ni atom through O-$p$ orbital. This O-$p$ mediated electron hopping between Ni-$e_g$ orbitals sitting on different atoms with opposite spin polarization renormalizes the one-electron level and shift the majority-spin Ni-$e_g$ orbitals to higher energy.  In PM phase (see Fig. \ref{fig_4} (d)), Ni-$e_{g}$ hybridization is also divergent, but their magnitude is weaker. The relatively weak electron hopping between Ni-$e_{g}$ orbital to another Ni-$e_{g}$ orbitals sitting on opposite-spin atom can be explained by Pauli-blocking. The spin fluctuation in Ni orbital in PM phase hampers electron hopping between Ni-$e_{g}$ orbitals sitting on neighboring sites due to Pauli-exclusion principles. This is consistent with the non-local screening mechanism suggested by NiO cluster model calculation ~\cite{kuo_epj_2017}.

% % in  Majority-spin electrons in an occupied Ni-$e_g$ hops to a unoccupied Ni-$e_g$ orbitals at the neighboring site via O-$p$ orbitals.
% This can be comfirmed by comparing density of states within charge self-consistent LDA+DMFT and \text{ab initio} LQSGW+DMFT. Fig. \ref{fig_3}(d) shows charge self-consistent LDA+DMFT results with U=XX and J=XX. If we make a comparision of Ni-$e_g$ hybridization function of charge self-consistent LDA+DMFT and \text{ab initio} LQSGW+DMFT, Both of them are strongest at an energy where O-p density of states dominates, but the magnitude of hybridiztion is much smaller and majorigy spin Ni-t$_{2g}$ and Ni-$e_g$ level splitting is smaller within charge self-consistent LDA+DMFT. Another way of confirming this idea is replacing Ni-$e_g$ hybridization within \textit{ab initio} LQSGW+DMFT with Ni-$e_g$ hybridization within charge self-consistent LDA+DMFT and plot the local Green's function. Fig. \ref{fig_3}(c) shows \textit{ab initio} DMFT local Green's function with charge self-consistent LDA+DMFT hybridization. This shows clearly that stonger electron hopping between Ni-$e_g$ orbitals gives rise to larger majorigy spin Ni-$e_g$ and Ni-$t_{2g}$ splitting.

Majority-spin Ni-$t_{2g}$-$e_{g}$ splitting as well as magnetic Ni-$t_2g$ splitting don't guarantee the two-peak structure but their splitting should match to each other for the distinct two-peak structure. Within charge self-consistent LDA+DMFT, there are non-negligible majority-spin Ni-$t_{2g}$-$e_{g}$ splitting as well as magnetic Ni-$t_{2g}$ splitting but the two-peak structure is not visible in the total density of states. Within LDA+DMFT, majority-spin Ni-$t_{2g}$-$e_{g}$ splitting is two times larger than the magnetic Ni-$t_{2g}$ splitting. Then minority-spin Ni-$t_{2g}$ peak hides the density dip between majority-spin Ni-$t_{2g}$ density and majority-spin Ni-$e_{g}$ density.

Non-negligible B peak intensity at the Neel temperature shown in Fig. 1(a) can be attributed to the short-range magnetic ordering. In real materials in their paramagnetic phase, there is short-range magnetic order, which has been neglected in our \textit{ab initio} calculation. Short range magnetic order can enhance both magnetic exchange splitting in $t_{2g}$ orbital and O-mediated electron hopping between Ni-$e_g$ orbitals in comparison to the ideal PM phase without shot-range magnetic order.

\textit{Conclusion.-} In conclusion, the enhancement of two-peak structure in valence photoemission experiments of NiO below Neel temperature is reproduced by using parameter-free \textit{ab intio} LQSGW+DMFT. The experimentally observed enhancement is triggered by the concerted effect of the magnetic exchange splitting of Ni-$t_{2g}$ orbitals caused by local Hund coupling and the majority-spin Ni $t_{2g}$-$e_{g}$ splitting governed by non-local intersite electron hopping. Our results provide an important example where the interplay between local and non-local physics generate a new signature in experiments. Our new understanding should be useful in designing advanced photovoltaic and spintronic devices based on the NiO.

\begin{acknowledgments}
  We thank G. Kotliar for fruitful discussion. B. Kang and S. Choi were supported by the U.S Department of Energy, Office of Science, Basic Energy Sciences as a part of the Computational Materials Science Program. This research used resources of the National Energy Research Scientific Computing Center (NERSC), a U.S. Department of Energy Office of Science User Facility operated under Contract No. DE-AC02-05CH11231.
\end{acknowledgments}

\bibliography{bktm}

\end{document}